\documentclass[12pt]{article}
\usepackage{epsf}

\def\lesssim{\mathrel{\hbox{\rlap{\hbox{\lower4pt\hbox{$\sim$}}}\hbox{$<$}}}}
\def\gtrsim{\mathrel{\hbox{\rlap{\hbox{\lower4pt\hbox{$\sim$}}}\hbox{$>$}}}}

\begin{document}
\baselineskip=24pt

\begin{centering}
\parindent 0mm

{\large\bf BACKWARD ASYMMETRY OF THE COMPTON SCATTERING BY AN
ISOTROPIC DISTRIBUTION OF RELATIVISTIC ELECTRONS: ASTROPHYSICAL
IMPLICATIONS}

\vspace{1cm}
{\bf S.Y. Sazonov, R.A. Sunyaev}

\vspace{1cm}
{\sl Max-Planck-Institut f\"ur Astrophysik, Garching 

Space Research Institute, Moscow
}

\end{centering}

\clearpage
{\parindent 0mm
The angular distribution of low-frequency radiation after single scattering
by an isotropic distribution of relativistic electrons considerably
differs from the Rayleigh angular function. In particular, the
scattering by an ensemble of ultra-relativistic electrons obeys the
law $p=1-\cos{\alpha}$, where $\alpha$ is the scattering angle; hence
photons are preferentially scattered backwards. We discuss some
consequences of this fact for astrophysical problems. We show that a
hot electron-scattering atmosphere is more reflective than a cold one:
the fraction of incident photons which become reflected having
suffered a single scattering event can be larger by up to 50 per
cent in the former case. This should affect the photon exchange
between cold accretion disks and hot coronae or ADAF flows in the
vicinity of relativistic compact objects, as well as the rate of
cooling (through multiple inverse-Compton scattering of seed photons
supplied from outside) of optically thick clouds of relativistic
electrons in compact radiosources. The backward scattering asymmetry
also causes spatial diffusion of photons to proceed slower in hot
plasma than in cold one, which is important for the shapes of
Comptonization spectra and time delays in the detection of soft
and hard radiation from variable X-ray sources.
}
\clearpage

\section{INTRODUCTION}

In our recent paper (Sazonov, Sunyaev, 1999a) we have called attention
to the fact that the differential cross-section for Thomson scattering
of low-frequency photons averaged over an isotropic distribution of
relativistic electrons is substantially different from the Rayleigh phase
function, which corresponds to the scattering by cold electrons. The
resulting angular function is backward-oriented, i.e. {\sl photons
tend to be scattered backwards}, rather than forwards (see Fig.~1). It
should be immediately emphasized that this angular function is a {\sl
characteristic of the scattering by an ensemble of electrons}. Similar
terminology is used when treating the Compton scattering in dense
plasmas, where collective effects are important (see, e.g., Bekefi, 1966).

The phenomenon mentioned above results from the combined operation of two
effects. One is that a photon is more likely to suffer a scattering from an
electron that is moving toward it, rather than away from it (the
probability of scattering is proportional to the factor
$1-\cos{\theta}\,v/c$, where $v$ is the electron velocity and $\theta$
is the angle through which the photon and electron encounter). The other
effect is that photons emerge after scattering preferentially in the
direction of the motion of the relativistic electron. The angular
function in the considered case contrasts with the forward-oriented
Klein-Nishina angular function which corresponds to the case of
scattering of energetic photons on a resting electron.

In the present paper, we consider some astrophysical implications of this 
peculiar scattering behaviour of hot plasma. The most evident 
consequence of the backward scattering law is that a larger (than in
the case of cold plasma) fraction of low-frequency photons incident on
the surface of an optically thick cloud of relativistic electrons will
be reflected from it after a single scattering by an electron. Therefore, less 
photons will participate in the Comptonization process (through multiple
scatterings) inside the cloud, and smaller will be the rate at which
the electrons lose their energy. We derive (in Section~3), using
classical results of the theory of radiative transfer in scattering
atmospheres, simple formulae for the albedos of atmospheres consisting
of either mildly or ultra relativistic thermal electrons. The results
obtained can be useful, in particular, for calculations of the photon
exchange between cold accretion disks and hot coronae (or outflows)
above them, or in the study of the illumination of an ADAF accretion
flow by external low-frequency radiation. 

Another consequence of the backward scattering asymmetry is that the
coefficient of spatial diffusion of photons in hot plasma 
should be different from that in the case of non-relativistic
plasma. This affects the shapes of Comptonization spectra and time
delays between the detections of soft and hard radiation from variable X-ray
sources. We discuss this issue in Section~4. 

It is necessary to mention that the phenomenon of anisotropic
(backward) scattering of low-frequency radiation in a hot plasma has 
been previously discussed in literature, and its various astrophysical
consequences have been studied in detail, mainly in the context of the
mechanisms of formation of hard X-ray spectra in compact sources
(Ghisellini et al., 1991; Titarchuk 1994; Stern et al., 1995; Poutanen,
Svensson, 1995; Gierlinski et al., 1997; Gierlinski et al., 1999). In
particular, the scattering angular function for an ensemble of
Maxwellian electrons was investigated in the paper by Haardt (1993),
where approximate formulae were obtained for it based on the results
of Monte-Carlo simulations. The main purpose of the present paper is
to present a set of exact analytical formulae for the scattering
angular function and some related quantities, which are applicable in the
ultra-relativistic and mildly relativistic limits.

\section{SCATTERING ANGULAR FUNCTION}

The scattering angular function $p(\mu_{\rm s})$ gives the probability of
scattering through a given angle $\alpha=\arccos{\mu_{\rm s}}$ of a
photon by an ensemble of electrons. Below $p(\mu_{\rm s})$ is
normalized so that the mean photon free path in the plasma is 
\begin{equation}
\lambda=\frac{1}{N_{\rm e}\sigma_{\rm T}\int p(\mu_{\rm s})\,d\mu_{\rm s}/2},
\label{lambda}
\end{equation}
where $N_{\rm e}$ is the electron number density and $\sigma_{\rm T}$ is the
Thomson scattering cross-section. Note that according to the
definition (\ref{lambda}), $\int p(\mu_{\rm s})\,d\mu_{\rm s}/2 < 1$ in the
general case (because of Klein-Nishina corrections).

We shall restrict ourselves in this paper to the case in which
low-frequency radiation is scattered by electrons that obey a
relativistic Maxwellian distribution. Two opposite limits within this
case can be considered: (A) of ultra-relativistic electrons, $kT_{\rm e}\gg
m_{\rm e}c^2$, and (B) of weakly relativistic electrons, $kT_{\rm
e}\ll m_{\rm e}c^2$ (where $T_{\rm e}$ is the electron
temperature). In case (A), the following asymptotic formula is
applicable (Sazonov, Sunyaev, 1999a):
\begin{equation}
p(\mu_{\rm s}) =1-\mu_{\rm s}.
\label{ang_limit}
\end{equation}
This expression exactly corresponds to the ideal case where
$T_{\rm e}\rightarrow\infty$ and $\nu\rightarrow 0$ ($\nu$ being the initial
photon energy). The angular function is backward-oriented in this case
(see Fig.~1). It is possible to add first-order correction terms, one
allowing for Klein-Nishina corrections and another taking into
account the finite energy of electrons, to equation (\ref{ang_limit})
(Sazonov, Sunyaev, 1999a): 
\begin{equation}
p(T_{\rm e},\nu,\mu_{\rm s}) =
1-\mu_{\rm s}-6\frac{h\nu}{m_{\rm e}c^2}\frac{kT_{\rm e}}{m_{\rm
e}c^2}(1-\mu_{\rm s})^2 +\frac{-1+3\ln{4}
+6\Gamma(0,m_{\rm e}c^2/kT_{\rm e})}{8}\left(\frac{m_{\rm
e}c^2}{kT_{\rm e}}\right)^2\mu_{\rm s}, 
\label{ang_maxwel1}
\end{equation}
where $\Gamma(\alpha,z)=\int_z^\infty x^{\alpha-1} e^{-x}dx$ is the
incomplete Gamma function. Formula (\ref{ang_maxwel1}) is a good
approximation if $kT_{\rm e}\gtrsim 2 m_{\rm e}c^2$, $h\nu kT_{\rm
e}\lesssim 0.01 (m_{\rm e}c^2)^2$.

In case (B), the angular function can be described by the formula
(Sazonov \& Sunyaev 1999b)
\begin{eqnarray}
p(T_{\rm e},\nu,\mu_{\rm s})=\frac{3}{4}\left[1+\mu_{\rm
s}^2-2(1-\mu_{\rm s})(1+\mu_{\rm s}^2) \frac{h\nu}{m_{\rm e}c^2}
+2(1-2\mu_{\rm s}-3\mu_{\rm s}^2+2\mu_{\rm s}^3)\frac{kT_{\rm e}}{m_{\rm e}c^2}
\right.
\nonumber\\
\left.
+(1-\mu_{\rm s})^2(4+3\mu_{\rm s}^2)\left(\frac{h\nu}{m_{\rm e}c^2}\right)^2
+(1-\mu_{\rm s})(-7+14\mu_{\rm s}+9\mu_{\rm s}^2-10\mu_{\rm
s}^3)\frac{h\nu}{m_{\rm e}c^2}\frac{kT_{\rm e}}{m_{\rm e}c^2}
\right.
\nonumber\\
\left.
+(-7+22\mu_{\rm s}+9\mu_{\rm s}^2-38\mu_{\rm s}^3+20\mu_{\rm
s}^4)\left(\frac{kT_{\rm e}}{m_{\rm e}c^2}\right)^2
+...\right],
\label{ang_maxwel2}
\end{eqnarray}
which is a good approximation when $kT_{\rm e}\lesssim 0.05 m_{\rm e}c^2$ and
$h\nu\lesssim 0.1 m_{\rm e}c^2$. The main term in the power series above,
$3(1+\mu_{\rm s}^2)/4$, is the usual Rayleigh function, which corresponds to
the non-relativistic case. 

Various examples of the angular function corresponding to the 
scattering on high-temperature electrons, as resulted from Monte-Carlo
simulations or calculated from equations (\ref{ang_maxwel1}) and
(\ref{ang_maxwel2}), are presented in Fig.~1.

\section{REFLECTION FROM A HOT PLANE-PARALLEL ATMOSPHERE}

The angular laws of scattering by ensembles of Maxwellian electrons
described above are readily applicable to any classical problem of
radiation transfer in scattering atmospheres. In fact, as long as the
evolution of the photon energy in time and space is not important,
i.e. the scattering angular function can be considered unchanging
(which takes place, e.g., if $h\nu$ remains so small that Klein-Nishina
corrections can be ignored), it is possible to apply some classical
results (found, e.g., in the textbooks by Chandrasekhar, 1950 and
Sobolev, 1963) to hot electron-scattering atmospheres just 
by substituting the ensemble-averaged angular function for the phase
function in the corresponding formulae. We shall consider below the
problem about the reflection of light from such an atmosphere.

Since Compton scattering involves the transfer of energy between
electrons and radiation, it is convenient to carry out the treatment
in terms of number of photons, rather than in terms of intensity. We
aim to determine what fraction of photons incident on the atmosphere
become reflected having suffered only one scattering event. At the
same time, that will immediately tell us what fraction of the incident
photons are capable of penetrating inside the atmosphere, undergoing
multiple scatterings and thus taking part in the Comptonization process (here
we are talking about an atmosphere with an optical depth $\tau\gtrsim
1$, otherwise a large fraction of photons will traverse the atmosphere
un-scattered). Throughout our discussion below, we restrict ourselves
to the case in which the incident photons are of sufficiently low
energy so that Klein-Nishina corrections are not important. It
should be specially noted that in this (Thomson) limit $\int
p(\mu_{\rm s})\,d\mu_{\rm s}/2=1$.

In the case of a semi-infinite atmosphere of optical depth
$\tau\rightarrow\infty$, the number of photons which have suffered a
single scattering process, emergent in a specified direction
$\mu=\cos{\theta}$, $\phi$ (relative to the normal to the
atmosphere), is given by the relation (Chandrasekhar, 1950)
\begin{equation}
N^{(1)}(\mu,\phi;\mu_0,\phi_0)=\frac{1}{4\pi(\mu+\mu_0)}
p(\mu,\phi;-\mu_0,\phi_0)F_0,
\end{equation}
where $\mu_0$, $\phi_0$ define the direction from
which the radiation is incident on the atmosphere, $F_0$ is the
incident photon flux per unit surface area, and $p$ is the scattering
phase function.

We may introduce the notion of atmosphere albedo with respect to
singly scattered photons:
\begin{equation}
A^{(1)}(\mu_0)=\frac{\int\int N^{(1)}(\mu,\phi;\mu_0,\phi_0)\mu\,d\mu
d\phi}{F_0},
\label{alb_def}
\end{equation}
which is a function of the incidence angle $\mu_0$.

The calculation in equation (\ref{alb_def}) is straightforward
for two scattering laws which are of interest to us: (A) $p=1-\mu_{\rm
s}$ and (B) $p=3(1+\mu_{\rm s}^2)/4$ (the Rayleigh law), for which the
single-scattering albedos are, respectively, 
\begin{eqnarray}
\mbox{(A)} A^{(1)}(\mu_0)= 
\frac{1}{2}\left[1+\frac{1}{2}\mu_0-\mu_0^2-\mu_0(1-\mu_0^2) 
\ln{\frac{1+\mu_0}{\mu_0}}\right],\\
\mbox{(B)} A^{(1)}(\mu_0)= 
\frac{3}{16}\left[\frac{8}{3}+\frac{1}{2}\mu_0-\mu_0^2-\frac{3}{2}
\mu_0^3+3\mu_0^4-\mu_0(3-2\mu_0^2+3\mu_0^4)\ln{\frac{1+\mu_0}{\mu_0}}\right].
\label{alb_lim}
\end{eqnarray}

Case (B) corresponds to a situation in which the electrons are
cold ($kT_{\rm e}\ll m_{\rm e}c^2$), while case (A) will be realized,
as follows from the results of the previous section, if the plasma is very
hot ($kT_{\rm e}\gg m_{\rm e}c^2$). The reflection properties of the atmosphere
that obeys the scattering law $p=1-\mu_{\rm s}$ are noticeably different from
those of the cold atmosphere, as demonstrated by Fig.~2(a). In
particular, the single-scattering albedo for normally incident photons
($\mu_0=1$) is 0.25 for the atmosphere consisting of
ultra-relativistic electrons, whereas the corresponding value for case
(B) is 0.168. This means that 50 per cent more photons are reflected,
having experienced a single scattering event, from the hot atmosphere
than from the cold one. It is evident from Fig.~2(a) that for
$kT_{\rm e}\gtrsim 2 m_{\rm e}c^2$ the albedo approaches the value
corresponding to the limiting, ultra-relativistic case (A). 

If the electron temperature is only mildly relativistic, the
appropriate expression for the albedo can be found by integrating the
angular function (\ref{ang_maxwel2}):
\begin{eqnarray} 
A^{(1)}(\mu_0)= 
\frac{3}{16}\left\{\frac{8}{3}+\frac{1}{2}\mu_0-\mu_0^2-\frac{3}{2}
\mu_0^3+3\mu_0^4-\mu_0(3-2\mu_0^2+3\mu_0^4)\ln{\frac{1+\mu_0}{\mu_0}}
+\left[-2\mu_0+6\mu_0^2
\right.\right.
\nonumber\\
\left.\left.
+16\mu_0^3-\frac{106}{3}\mu_0^4-10\mu_0^5
+20\mu_0^6-\mu_0(-2+16\mu_0^2-42\mu_0^4+20\mu_0^6)\ln{\frac{1+\mu_0}{\mu_0}}
\right]\frac{kT_{\rm e}}{m_{\rm e}c^2}
\right\}, 
\label{alb_mild}
\end{eqnarray}
where we have quoted only the correction term proportional to
$kT_{\rm e}/m_{\rm e}c^2$. This formula is a good approximation if
$kT_{\rm e}\lesssim 0.2 m_{\rm e}c^2$. As follows from both equation
(\ref{alb_mild}) and Fig.~2a, the single-scattering albedo of the
atmosphere consisting of mildly relativistic electrons is only
slightly larger than that of the atmosphere with the Rayleigh
scattering law. The relative difference is less than 5 per cent if
$kT_{\rm e}\lesssim 0.1 m_{\rm e}c^2$. We may conclude that in most
practical applications, it should be sufficiently accurate to use the
albedo corresponding to the cold atmosphere when reflection of
radiation from plasmas with $kT_{\rm e}\lesssim 0.1 m_{\rm e}c^2$ is
investigated.

Although formulae (\ref{alb_lim}) and (\ref{alb_mild}) are, strictly,
applicable only to an atmosphere whose optical thickness is infinite, a
similar increase in the albedo during the transition from the cold to hot
electrons occurs in the case where the atmosphere is transparent. As
follows from results of Monte-Carlo simulations, the single-scattering
albedo with respect to normally incident photons is increased by
approximately 50 per cent on going from the cold to ultra-relativistic
electrons for arbitrary values of the atmosphere optical depth. An
example in which $\tau=0.5$, a value typical for the advection flows
near accreting compact objects, is presented in Fig.~2(b).

We can also calculate the single-scattering albedo of an optically thick
atmosphere for the case of isotropic incident radiation (when
$I_0(\mu_0,\phi_0)=const$):
\begin{eqnarray}
\mbox{(A)}~~&~& A^{(1)}_{\rm iso}=\frac{7-4\ln{2}}{15}\approx 0.282,\\
\mbox{(B)}~~&~& A^{(1)}_{\rm iso}=\frac{26-27\ln{2}}{35}\approx 0.208.
\end{eqnarray}
We obtain the result that 36 per cent more photons are reflected (after a
single scattering) from the ultra-hot atmosphere than from the cold one. 

\section{PHOTON DISTRIBUTION OVER THE ESCAPE TIME FROM A CLOUD OF\\ ELECTRONS}

In the previous section we showed that the fraction of low-frequency photons
incident on a cloud of electrons which become the seed photons for the
Comptonization inside the cloud, i.e. the normalization of the
distribution of multiply (more than once) scattered photons over the
number of scatterings, depends on the plasma temperature. In
this section, we shall draw our attention to the shape of this distribution, 
which is also affected, but to a lesser degree, by the scattering
angular function.

Sunyaev, Titarchuk (1980) and Payne (1980) obtained analytical
formulae describing the distribution of photons over the time of their
escape from a spherical, optically thick plasma cloud. The calculation
of these authors, carried out in the diffusion approximation,
implicitly assumed that the probabilities are equal for the 
scatterings in the forward ($\mu_{\rm s}>0$) and backward ($\mu_{\rm s}<0$)
directions. This condition is, for example, indeed met when the
scattering angular function is spherical ($p=1$) or Rayleigh
($p=3(1+\mu_{\rm s}^2)/4$). The latter opportunity is the case when both the
photons and the electrons are non-relativistic ($h\nu,kT_{\rm e}\ll
m_{\rm e}c^2$). However, as we showed in Section~2, the forward-backward
symmetry is violated if the electrons are substantially relativistic.

It is easy to modify the solution of Sunyaev, Titarchuk (1980) so
that it would take into account the dependence of the scattering law
on plasma temperature. To this end we should use the correct value of
the diffusion coefficient, which is given by (see, e.g., Weinberg,
Wigner, 1958)
\begin{equation}
D=\frac{1}{3(\sigma/\sigma_{\rm T})(1-\langle\mu_{\rm s}\rangle)},
\label{dif_coef}
\end{equation}
where $\sigma$ is the total scattering cross-section, and 
\begin{equation}
\langle\mu_{\rm s}\rangle=\frac{\int\mu_{\rm s} p(\mu_{\rm
s})\,d\mu_{\rm s}/2}{\int p(\mu_{\rm s})\,d\mu_{\rm s}/2}.
\label{mu_aver}
\end{equation}
In the non-relativistic limit, $\sigma=\sigma_{\rm T}$,
$\langle\mu_{\rm s}\rangle=0$ and $D=1/3$, and it is the latter value
which was used in (Sunyaev, Titarchuk, 1980). 

We shall assume here (as we did in Section~3) that the energy of a
photon remains small as it diffuses out of the cloud suffering
multiple scatterings, hence Klein-Nishina corrections are not
important. In this case $\sigma=\sigma_{\rm T}$, and (see
eq. [\ref{dif_coef}]) only the dependence of $\langle\mu_{\rm s}\rangle$ on
the scattering law $p(\mu_{\rm s})$ determines the magnitude of the
diffusion coeffient. 

In the ultra-relativistic case ($kT_{\rm e}\gg m_{\rm e}c^2$), is can
readily be found, using equation (\ref{ang_maxwel1}), that
$\langle\mu_{\rm s}\rangle=-1/3$ and, consequently,
\begin{equation}
D=\frac{1}{4}.
\label{dif_ultra}
\end{equation}
Therefore, diffusion of low-frequency photons proceeds slower, by a
factor of $4/3$, in ultra-relativistic plasma than in cold
one. The value (\ref{dif_ultra}), however, has little practical
significance. Indeed, a photon acquires a huge amount of energy upon
scattering from an ultra-relativistic electron of energy $\gamma
m_{\rm e}c^2$ ($\nu^\prime/\nu\sim \gamma^2$), hence it typically becomes
relativistic (being originally low-frequency) after few
scatterings and our Thomson-limit set-up of the problem becomes invalid.

In the mildly relativistic limit, we derive, using equation
(\ref{ang_maxwel2}), that $\langle\mu_{\rm s}\rangle\approx -2kT_{\rm
e}/5m_{\rm e}c^2$ and
\begin{equation}
D(T_{\rm e})=\frac{1}{3(1+2kT_{\rm e}/5m_{\rm e}c^2)}.
\label{dif_mild}
\end{equation}

We would like to point out that our treatment above is similar to that
in (Illarionov et al., 1979; Grebenev, Sunyaev 1987), where the
diffusion coefficient that corresponds to the case of scattering of
relativistic photons by cold electrons was derived. There have also
been investigations (Cooper, 1974; Shestakov et al., 1988)
considering the effect of relativistic corrections on the so-called
transport cross-section, which is used to describe the spatial
diffusion of radiation in terms of intensity, rather than in terms of
number of photons, as we do here.

Using the diffusion coefficent in the form (\ref{dif_mild}), it is easy to
recalculate distribution functions over the photon escape time
that were found by Sunyaev and Titarchuk (1980) for various
distributions of the sources of photons over a spherical cloud. For
example, if the source is situated at the center of the cloud, and we
are interested in the fate of photons that undergo more scatterings
than the average number, i.e. the dimensionless time $u\equiv
c\sigma_{\rm T} N_{\rm e}t\gg u_0=0.5\tau^2$ (where $\tau\gg 1$ is the
optical depth of the cloud along its radius), then

\begin{equation}
P(u)=\frac{2\pi^2 D(T_{\rm e})}{(\tau+2D(T_{\rm e}))^2}
\exp{\left[-\frac{u\pi^2 D(T_{\rm e})}{(\tau+2D(T_{\rm e}))^2}\right]},
\label{p_u}
\end{equation}
where $D(T_{\rm e})$ is given by equation (\ref{dif_mild}).

In Fig.~3, we have plotted $P(u)$ for a number of values of
$kT_{\rm e}$. We see that the agreement between Monte-Carlo results and the
result of equation (\ref{p_u}) is fairly good up to $kT_{\rm e}\sim 0.4
m_{\rm e}c^2$. The temperature correction to the diffusion coefficient
causes photons to spend more time (and suffer more inverse-Compton
scatterings) in the cloud.

The knowledge of how photons are distributed over the time they spend
in a scattering cloud is required in virtually all problems related to
Comptonization in astrophysical plasmas. In particular, the shapes of
Comptonization spectra and the time delays between the soft and hard
components of radiation coming from variable X-ray sources strongly
depend on this quantity. To solve such problems, it is necessary to consider a
related question about the evolution of the photon energy 
with time, which is beyond the scope of this letter. We note, however,
that much work in this direction (in connection to relativistic
thermal plasmas) has already been done (Titarchuk, 1994; Hua,
Titarchuk, 1996; see also references therein).

This research has been supported in part by the Russian Foundation for Basic
Research through grant 97-02-16264. The authors are grateful to the 
referee, Sergei Grebenev, for valuable comments. We would also like to
thank Drs A.A. Zdziarski and  C. Done, who pointed out the
necessity of mentioning in the paper a number of published works 
that address questions which are closely related to the
subject of the present study.

\clearpage
\centerline{REFERENCES}
\parindent 0mm

Bekefi G.// Radiation Processes in Plasmas. New York. Wiley, 1966.

Chandrasekhar S.// Radiative Transfer. New York. Dover, 1950.

Cooper G.// J. Quant. Spectr. Rad. Transf., 1974, v. 14, p. 887.

Ghisellini G., George I.M., Fabian A.C., Done C.//
Mon. Not. R. Astron. Soc., v. 248, p. 14, 1991.

Gierlinski M., Zdziarski A.A., Done C., Johnson W., Ebisawa K., Ueda
Y., Haardt F.//  Mon. Not. R. Astron. Soc., v. 288, p. 958, 1997.

Gierlinski M., Zdziarski A.A., Poutanen J., Coppi P.S., Ebisawa K.,
Johnson W.N.// Mon. Not. R. Astron. Soc., v. 309, p. 496, 1999.

Grebenev S.A., Sunyaev R.A.// Sov. Astron. Lett., 1987, v. 13, p. 438.

Haardt F.// Astrophys. J., v. 413, p. 680, 1993.

Hua X.-M., Titarchuk L.G.// Astrophys. J., v. 469, p. 280, 1996.

Illarionov A., Kallman T., McCray R., Ross R.// Astrophys. J.,
v. 228, p. 279, 1979.

Payne D.G.// Astrophys. J., 1980, v. 237, p. 951.

Pozdnyakov L.A., Sobol I.M., Sunyaev R.A.// Astrophys. and Space
Phys. Rev., 1983, v. 2, p. 189 (ed. Sunyaev. Chur. Harwood Academic
Publishers).

Poutanen J., Svensson R.// Astrophys. J., v. 470, p. 249, 1996.

Sazonov S.Y., Sunyaev R.A.// Astron. Astrophys., v. 354, L53, 2000.  

Sazonov S.Y., Sunyaev R.A.// Astrophys. J. (submitted);
astro-ph/9910280, 15 Oct., 1999b.

Shestakov A.I., Kershaw D.S., Prasad M.K.//
J. Quant. Spectr. Rad. Transf., v. 40, p. 577, 1988.

Sobolev V.V.// A Treatise on Radiative Transfer. Princeton. Van
Nostrand, 1963.

Stern B.E., Poutanen J., Svensson R., Sikora M., Begelman M.C.//
Astrophys. J., v. 449, L13, 1995.

Sunyaev R.A., Titarchuk L.G.// Astron. Astrophys., v. 86, p. 121, 1980.

Titarchuk L.G.// Astrophys. J., v. 434, p. 570, 1994.

Weinberg A.M., Wigner E.P.// The Physical Theory of Neutron Chain
Reactors. Chicago. University Chicago Press, 1958.

\begin{figure}
\epsfxsize=14.5cm
\epsffile[90 275 560 595]{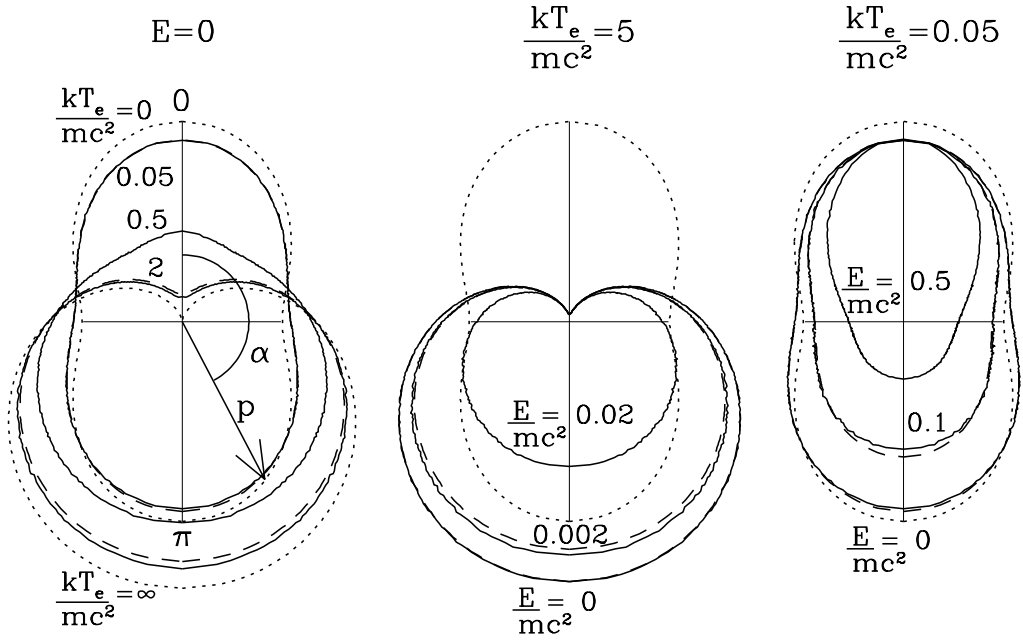}
\caption{Angular function (in polar coordinates) for the Compton
scattering of photons of energy $E$ by an ensemble of relativistic
Maxwellian electrons with temperature $T_e$. Solid lines show
Monte-Carlo simulation results. Dashed lines represent approximations
(when available) given either by Eq.~(\ref{ang_maxwel1}) or by
Eq.~(\ref{ang_maxwel2}). The dotted lines represent two extreme cases,
one of vanishing temperature, which corresponds to the Rayleigh
angular function, and the other of ultra-relativistic electrons
($kT_e\gg m_ec^2$). In the latter case (presented only in the left
panel), the scattering obeys the law $p=1-\cos{\alpha}$.
\label{fig1}}
\end{figure}

\begin{figure}
\epsfxsize=14.5cm
\epsffile[45 170 560 680]{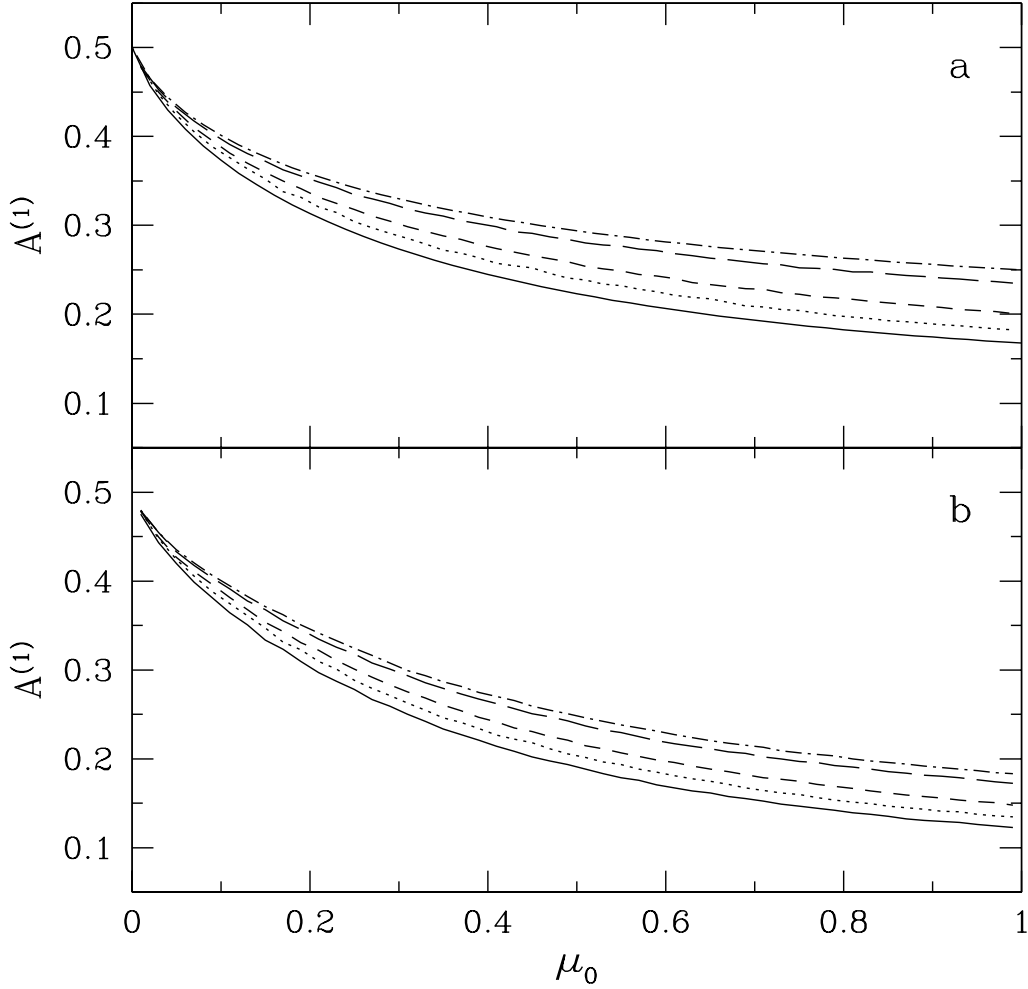}
\caption{Albedo of an electron-scattering atmosphere with respect to
singly scattered photons as a function of the incidence angle
$\alpha=\arccos{\mu_0}$. The radiation is of low frequency, $h\nu kT_e\ll
(m_ec^2)^2$. Monte-Carlo simulation results are shown. (a) The optical
thickness of the atmosphere is infinite ($\tau\rightarrow\infty$). Different
curves correspond to various electron temperatures: $kT_e\ll m_ec^2$
(solid line), $0.2 m_ec^2$ (dotted line), $0.5 m_ec^2$ 
(short-dashed line), $2 m_ec^2$ (long-dashed line) and $\gg m_ec^2$
(dash-dotted line). (b) The same as (a), but for a transparent
atmosphere with $\tau=0.5$.
\label{fig2}}
\end{figure}

\begin{figure}
\epsfxsize=14.5cm
\epsffile[45 170 560 680]{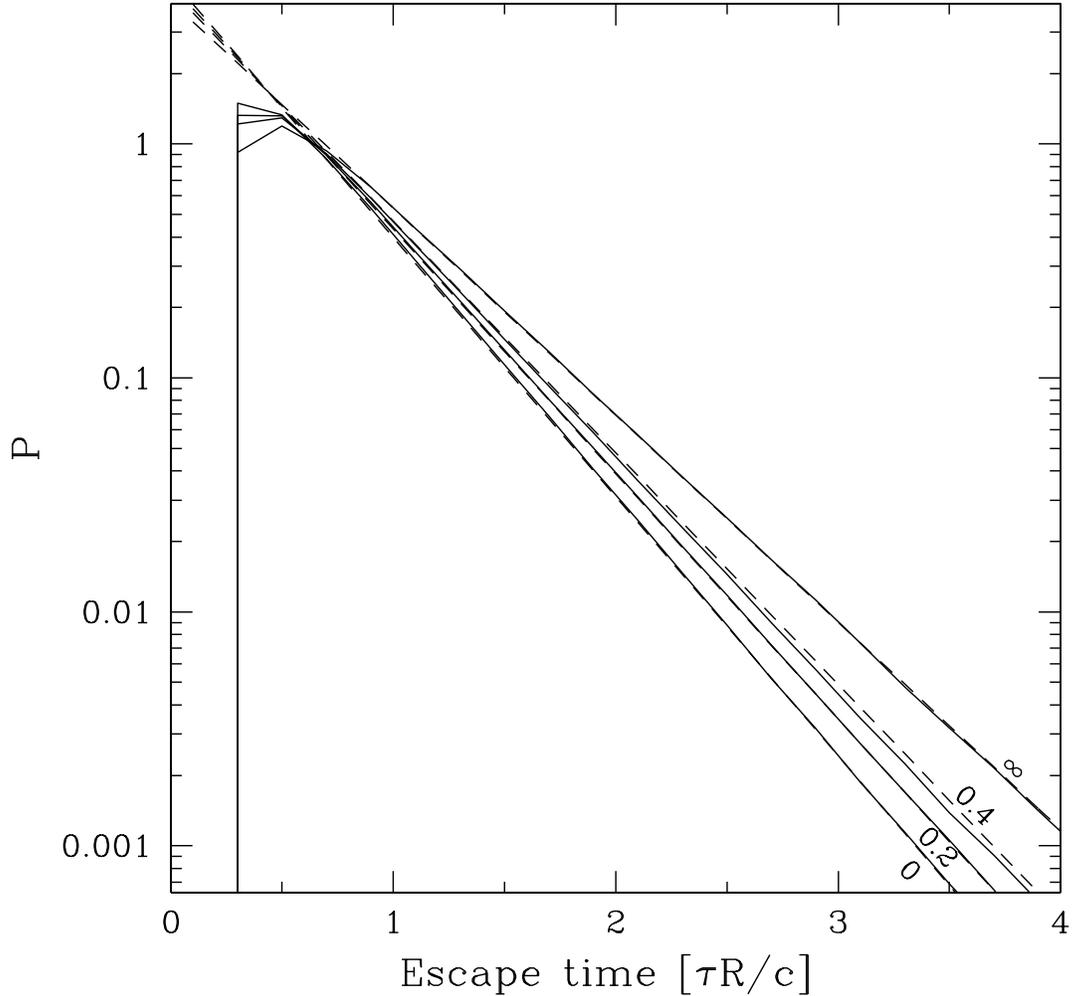}
\caption{Distribution of low-frequency photons over the time of
their escape from a cloud of electrons with an optical depth of $\tau=5$
($R$ is the radius of the cloud), for various plasma temperatures
(curves are labelled with the corresponding values of
$kT_e/m_ec^2$). The solid lines are Monte-Carlo simulation
results. The dashed lines show the results of the calculation by the 
approximate formula (\ref{p_u}), which can be used for describing the
exponential (power-law in the linear-logarithmic coordinates used)
part of the distribution function. The diffusion coeffient $D$
appearing in equation (\ref{p_u}) is given by Eq. (\ref{dif_ultra}) in
the case $kT_e\gg m_ec^2$ and by Eq. (\ref{dif_mild}) in the case
$kT_e\lesssim 0.4 m_ec^2$.
\label{fig3}}
\end{figure}

\end{document}